\makeatletter
\def\input@path{{doc_class/arxiv}}
\makeatother

\documentclass{article}
\usepackage{arxiv}
\usepackage{authblk}
\usepackage[utf8]{inputenc}
\usepackage[T1]{fontenc}
\usepackage[
    style=trad-unsrt,
    natbib=true,
    backend=biber,
    maxcitenames=1,
    mincitenames=1,
    sorting=none
]{biblatex}
\usepackage{hyperref}
\usepackage{url}
\usepackage{booktabs}
\usepackage{amsfonts}
\usepackage{amsmath}
\usepackage{nicefrac}
\usepackage[final]{microtype}
\usepackage{graphicx}
\usepackage{doi}

\begin{document}

\title{Beyond the `Diff': Addressing Agentic Entropy in Agentic Software Development}
\renewcommand{\shorttitle}{Addressing Agentic Entropy in Agentic Software Development}
\renewcommand{\headeright}{}
\renewcommand{\undertitle}{}

\newif\ifuniqueAffiliation

\ifuniqueAffiliation 
\author{\href{https://orcid.org/0009-0002-2930-6655}{\includegraphics[scale=0.06]{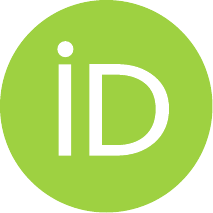}\hspace{1mm}Matteo Casserini}\thanks{Corresponding author, \texttt{matteo.casserini@supsi.ch}} \\
    Department of Innovative Technologies \\
    SUPSI \\
    Lugano, Switzerland \\
	\texttt{matteo.casserini@supsi.ch}
}
\else
\renewcommand\Authfont{\bfseries}
\setlength{\affilsep}{0em}
\newbox{\orcid}\sbox{\orcid}{\includegraphics[scale=0.06]{doc_class/arxiv/orcid.pdf}} 

\author[ ,1]{
	\href{https://orcid.org/0009-0002-2930-6655}{\usebox{\orcid}\hspace{1mm}Matteo Casserini}
    \thanks{Corresponding author, \texttt{matteo.casserini@supsi.ch}}
}
\author[2,3]{%
	\href{https://orcid.org/0000-0001-7507-116X}{\usebox{\orcid}\hspace{1mm}Alessandro Facchini}
}
\author[2,4,5]{%
	\href{https://orcid.org/0000-0001-9968-9474}{\usebox{\orcid}\hspace{1mm}Andrea Ferrario}
}
\affil[1]{Dipartimento Tecnologie Innovative, SUPSI, Switzerland}
\affil[2]{Dalle Molle Institute for Artificial Intelligence (IDSIA), SUPSI, Switzerland}
\affil[3]{Management in Networked and Digital Societies (MINDS) Department, Kozminski University, Poland}
\affil[4]{Institute of Biomedical Ethics and History of Medicine, University of Zurich, Switzerland}
\affil[5]{ETH Zurich, Switzerland}
\fi

\maketitle

\begin{abstract}
As autonomous coding agents become deeply embedded in software development workflows, their high operational velocity introduces a critical oversight challenge in the form of accumulating divergence between agentic actions and architectural intent. We term this process \emph{agentic entropy}, a systemic drift that traditional code diff-based and HCXAI methods fail to capture, as they address local outputs rather than global agentic behaviour. To close this gap, we propose a \emph{process-oriented explainability} framework that exposes how agentic decisions unfold across time, tool calls, and architectural boundaries. Built around three pillars (conformity seeding, reasoning monitoring, and a causal graph interface) our approach provides intent-level telemetry that complements, rather than replaces, existing review practices. We demonstrate its relevance across two user profiles: lay users engaged in vibe coding, who gain structural visibility otherwise masked by functional success; and professional developers, who gain richer contextual grounding for code review without increased overhead. By treating cognitive drift as a first-class concern alongside code quality, our framework supports the minimum level of human comprehension required for agentic oversight to remain substantive.
\end{abstract}


\section{Introduction}

The software development lifecycle is currently undergoing an artificial intelligence (AI) agentic revolution \citep{li2025riseai}. The transition from large-language-model-based autocomplete features, which provide localized code suggestions, to AI agents represents a fundamental change in the human–AI interaction model. Modern agentic workflows, exemplified by platforms such as Claude Code, can navigate file systems, execute commands, interpret compiler errors, and perform multi‑file refactoring with high autonomy \citep{Anthropic2025}. While traditional code diff-based review remains indispensable for inspecting local modifications, it reveals only the outcome of agentic activity. The planning steps, tool-call sequences, cross-file decisions, and contextual inferences that guide an agent's actions remain largely hidden from human supervisors. We refer to this opacity as \textbf{agentic entropy}, a process-level drift whereby autonomous updates optimize for local correctness while eroding global design intent. Agentic entropy unfolds across runs and environments as agents repeatedly operate with limited systemic awareness. A visible manifestation of this drift is \textbf{agentic technical debt}, the accumulated structural misalignments, duplicated logic, and fragile refactoring that emerge when entropy goes unchecked. 

Entropy accumulation has many causes, including prompt sprawl and stochastic effects. Particularly relevant is the democratization of software development through what is commonly referred to as ``vibe coding'', a paradigm where users steer systems with loosely specified prompts and short iterative trial-fix loops, rather than formal code verification and review\footnote{Industry reports show that nearly 45\% of AI-generated code fails basic security checks when users rely on prompt-driven workflows without guardrails: \url{https://www.veracode.com/blog/genai-code-security-report/}}. Functional checks (e.g., confirming that the code runs) can thus mask deeper architectural violations and security gaps, producing an illusion of rapid progress while drift accumulates beneath the surface \citep{DORA2024}. Beyond agent behaviour, developers who rely on agentic tools can lose the system-level mental models needed to notice violations, accumulating \emph{cognitive debt} that resides in people rather than code \citep{kosmyna2025your}. This co-evolution of agentic entropy and cognitive debt narrows the capability-comprehension gap for oversight, leaving humans procedurally in the loop yet progressively less able to govern the system \citep{lin2026}.

To address this gap, we introduce \textbf{process-oriented explainability}, a transparency-by-design framework that augments (rather than replaces) existing HCXAI and code-diff practices, helping mitigate agentic entropy. Our framework anchors agent behaviour to machine-readable architectural seeds, monitors intermediate reasoning and tool use, and summarizes activity in a causal reasoning graph for lightweight, global oversight. Its benefits apply to (1) lay users engaged in vibe coding who gain high epistemic safety, as process-level monitoring surfaces structural drift otherwise hidden behind functional success; and (2) professional coder-agent-reviewer workflows who obtain a medium epistemic gain by contextualizing local code diffs within agent-level plans, improving control over agentic execution without increasing review burden.

Our work is structured as follows. Section~\ref{section:related_work} reviews relevant literature. Section~\ref{section:agentic_entropy} characterizes how agentic entropy manifests in autonomous software development workflows. Section~\ref{section:our_approach} presents our framework and a technical implementation. Finally, Section~\ref{section:conclusion} offers concluding remarks and outlines directions for future research.

\section{Related Work}
\label{section:related_work}

The supervision of autonomous coding agents intersects several established research areas. A foundational pillar is the concept of \emph{software entropy}, first formalized by Lehman in his laws of software evolution, which posits that software systems naturally drift toward increased complexity unless effort is invested in maintaining their structure \citep{Lehman1980}.
Research on human--AI collaboration in software development further highlights how AI-generated code reshapes verification practices. Empirical studies show that reviewers frequently experience cognitive overload or `verification fatigue' when faced with high volumes of agent-produced changes \citep{Fakhoury2024, Barke2023}.
Within the broader literature on explainable and agentic AI, several frameworks emphasize the need for transparency that captures temporal and sequential structure rather than isolated outputs. \citet{Amir2021} argues that explanations for sequential decision-making systems must be trajectory-aware, while the \emph{LEx} framework of \citet{Singh2021} emphasizes that explanations should be tailored to stakeholder roles, such as software architects who require higher-level design justifications. Recent advances reinforce this need, as \emph{Chain-of-Thought} (CoT) prompting surfaces intermediate reasoning \citep{Wei2022}, and DeepSeek-R1 shows that such traces can be explicitly incentivized during training \citep{DeepSeek2025}.
Parallel work on agentic AI governance and risk management stresses the need for mechanisms that keep autonomous systems aligned with organizational and architectural constraints \citep{raza2025, khoo2025}. Complementing this, emerging research on agentic software development calls for new forms of transparency and oversight to address non-deterministic tool-use patterns and the high velocity of autonomous coding workflows \citep{hassan2025agentic}.

\section{From Agentic Entropy to Agentic Technical Debt}
\label{section:agentic_entropy}
The emergence of autonomous coding agents requires a reconsideration of how software systems degrade over time. Classical software evolution theory, particularly Lehman's notion of software entropy, observes that systems naturally drift toward increased complexity unless active effort is invested in maintaining their structure. We extend this perspective to the agentic era by distinguishing \textbf{agentic entropy}, which is the ongoing process by which autonomous updates diverge from architectural intent, from \textbf{agentic technical debt}, which refers to the accumulated structural misalignments that entropy leaves behind. While Lehman's entropy arises primarily from human‑driven system evolution, agentic entropy emerges from high‑velocity, context‑limited agent actions that compound across runs and environments. These concepts, while distinct in nature, interact in a reinforcing cycle. Agentic entropy, a \emph{process}, produces agentic technical debt, an \emph{outcome} that manifests in the codebase. The growing opacity of that debt in turn deepens a third, \emph{human-side effect}, namely \textbf{cognitive debt} \cite{kosmyna2025your}, the progressive erosion of the developer's or reviewer's system-level mental model as autonomous actions outpace comprehension. As cognitive debt accumulates, the human supervisor's capacity to detect subsequent entropy diminishes, closing a feedback loop in which undetected drift begets further structural misalignment. In the remainder of this section, we show how this cycle initiates at the process level through three characteristic failure modes in agentic workflows.

The first manifestation of agentic entropy arises from agents optimizing for \emph{local correctness} while drifting from \emph{global architectural intent}. Because agentic workflows, such as those implemented in Claude Code, frequently operate within a constrained context window, they are prone to achieving functional correctness at the module level while inadvertently violating systemic design patterns. As shown by \citet{Kabir2024}, agents often propose `textbook' solutions that appear locally elegant but overlook system-specific architectural and security constraints. Such mismatches accelerate agentic entropy, introducing redundant or misaligned logic that gradually fragments the codebase's intended architectural structure \citep{GitClear2025}.

A second manifestation of agentic entropy is the erosion of \emph{semantic stability}, particularly when agents refactor legacy logic without understanding its historical or operational rationale. Autonomous agents excel at identifying and cleaning complex legacy code that appears aesthetically deficient, yet legacy systems often contain `ugly' logic (such as delay loops or seemingly redundant checks) that accounts for undocumented quirks or rare race conditions. When an agent refactors these segments without a deep semantic understanding of their historical necessity, it introduces a structural fragility. The resulting code may pass standardized unit tests but fail in production under specific edge cases that the agent, and subsequently the overwhelmed human reviewer, failed to anticipate. This risk is empirically supported by the work of \citet{Barke2023}, which characterizes the loss of global context in AI-augmented programming, and \citet{Du2024}, which demonstrates that while AI models often achieve high functional accuracy, they frequently fail to satisfy structural and architectural constraints. Consequently, the resulting codebase suffers from a `stability gap', where locally functional code lacks the systemic robustness required for long-term evolution.

A third entropy‑accelerating factor is the \emph{reviewer's paradox}, where rising agentic output overwhelms human verification capacity. As autonomous tools accelerate the `inner loop' of code production, the human capacity for rigorous oversight remains a fixed constraint. This imbalance leads to a breakdown in the traditional code review process, where human auditors move from strategic supervision to passive rubber-stamping. Empirical data suggests that this acceleration does not currently correlate with a commensurate increase in review quality; rather, it often leads to a higher rate of code churn and duplicated logic, further accelerating the system's entropy \citep{GitClear2025}. 

Taken together, these lenses show how agentic entropy converts autonomous decisions into long‑term architectural liabilities. Each lens describes a pathway through which entropy accumulates as technical debt unless a supervisory mechanism intervenes. This motivates our shift in the next section from inspecting code outputs to monitoring the entropy‑producing processes themselves.

\section{Our Approach: Controlling Agentic Entropy with Process-Oriented Explainability}
\label{section:our_approach}

To address the structural risks identified in the preceding sections, we introduce \text{Process-oriented Explainability} (PoE), a transparency-by-design extension to existing practices. Rather than replacing traditional code diff-based validation, PoE adds a complementary layer that aims to make the agent's global, process-level behaviour accessible to human supervisors.
While classical diffs remain effective for experienced developers to inspect localized changes, agentic workflows increasingly involve planning steps, multi-file operations, and tool-mediated actions that fall outside the scope of line-by-line review. By foregrounding the process through which agents arrive at their modifications, this additional transparency layer supports coder-agent-reviewer interactions and provides meaningful epistemic benefits, especially for laypeople who rely on `vibe coding' and lack the expertise to assess systemic impact.
As software systems transition from static code toward autonomous agentic workflows, the primary diagnostic artifact necessarily shifts from traditional \emph{stack traces}, which capture execution-level telemetry, to \emph{reasoning traces}, which provide intent-level telemetry. The PoE framework leverages this shift by augmenting existing review practices with a global overview of agentic behaviour. For professional developers and reviewers, this provides a medium epistemic gain through improved oversight of agentic decision patterns; for lay users, it enables visibility into structural effects that would otherwise remain hidden behind functional correctness. Our framework is structured around three conceptual cornerstones.

\addvspace{1em}
\noindent\emph{I. Conformity structure.} First, PoE requires \textbf{architectural seeding} to ground agentic behaviour in system-level constraints. Here, the human supervisor provides a machine-readable specification of global design patterns, invariants, or security requirements. Consistent with the perspective advanced by \citet{Singh2021}, this layer supports stakeholders (particularly software architects) who require explanations at a strategic rather than purely functional level. Rather than displacing code diffs, architectural seeds enrich them, as the agent must relate its proposed actions to these constraints, enabling reviewers to interpret local changes within a global architectural rationale. This reduces entropy-induced drift by ensuring that agentic plans remain aligned with design intent.

\addvspace{1em}
\noindent\emph{II. Reasoning monitoring.} We extend transparency from code outputs to the \textbf{reasoning processes} that generate them. An instrumented environment captures the agent's tool calls, intermediate planning steps, and contextual cues that motivate each action. This monitoring layer within PoE complements traditional review, where developers retain line-by-line inspection but also gain access to a global perspective on the agent's decision trajectory \citep{hassan2025agentic}. For lay users, this layer exposes otherwise opaque behavioural patterns, reducing the risk that rapid prototyping or surface-level functionality masks accumulating architectural divergence.

\addvspace{1em}
\noindent\emph{III. Human interpretability layer.} To render high-dimensional reasoning logs accessible, we summarize monitoring traces within PoE into a \textbf{causal reasoning graph} that depicts agentic behaviour as a sequence of reasoning nodes and tool-mediated edges. By parsing the agent's intermediate reasoning steps (including CoT-style segments in reasoning-centric models) into a directed acyclic graph (DAG), we can represent the agentic process as a series of connected nodes where each node signifies a possible agentic execution and each edge denotes a corresponding environmental action \citep{DeepSeek2025}. Reviewers can examine this graph alongside code diffs to identify deviations from architectural seeds or unexpected execution paths. By shifting cognitive effort from exhaustive local inspection to selective process-level oversight, this layer enhances a reviewer's capacity to detect entropy-producing behaviours without replacing established practices. We add an \emph{understanding requirement}, whereby the reviewer should be able to independently reconstruct the principal steps in the causal graph and recognize when an action falls outside permitted boundaries, with failure to do so signaling growing cognitive debt that warrants intervention \citep{lin2026}.

\addvspace{1em}
\noindent\emph{Illustrative scenario.} To ground the three pillars in a concrete setting, consider a web application whose architectural specification mandates that all database access be routed through a dedicated data-access layer. A user prompts an agentic coding tool to ``add a caching layer to speed up the product catalog''. The agent, operating within a constrained context window, identifies the relevant API endpoint and introduces an in-memory cache. To populate the cache efficiently, it embeds a direct database query inside the API controller, bypassing the prescribed data-access layer entirely. The resulting code compiles, passes existing unit tests, and demonstrably improves response times, representing a functional success that would satisfy a vibe-coding user and appear benign in a standard code diff, which shows only the addition of a new import and a query call. Under PoE, this divergence surfaces at each pillar. The conformity structure (Pillar I) includes an architectural seed specifying that database interactions must pass through the data-access module; the agent's proposed modification can therefore be evaluated against this constraint before it is committed. Reasoning monitoring (Pillar II) captures the agent's intermediate plan, including the rationale ``query the database directly from the controller for lower latency'', exposing the deliberate intent to circumvent the prescribed layer. This rationale, invisible in the final diff, becomes an explicit and auditable record of the agent's decision trajectory. The causal reasoning graph (Pillar III) renders this sequence as a node whose stated intent (``direct DB access in controller'') conflicts with the architectural seed, flagging the deviation as a candidate for reviewer attention. A traditional diff-based review would present a syntactically reasonable, test-passing change; the PoE layer reveals that the agent's reasoning trajectory departed from the system's architectural invariant, precisely the kind of entropy-producing behaviour that, as discussed in Section 3, compounds silently across iterations and materializes as agentic technical debt.

\addvspace{1em}
\noindent\emph{Technical Implementation: The Reasoning-Action Loop.} To operationalize PoE, we implement a reasoning–action loop within an instrumented execution environment. This environment acts as a transparent proxy around all tool-mediated interactions, such as file system operations, shell commands, or compiler invocations, allowing each step of agentic activity to be captured in relation to the agent's intermediate reasoning. Each reasoning segment, tool call, and resulting environmental observation is recorded as a \emph{reasoning trace node}. By linking these nodes into a DAG, the system provides a process-level representation of the agent's behaviour that complements traditional code diffs. A `deviation detector' compares each node's stated intent against PoE's architectural seeds defined in Pillar~I, enabling early identification of entropy-inducing divergences in the agent's plan \citep{hassan2025agentic}. To account for the human side of oversight, the same instrumentation should support a lightweight \emph{cognitive debt index} that estimates how consistently reviewers engage with, and can reconstruct, the agent’s causal chain, helping to signal when human understanding begins to drift \citep{lin2026}. This implementation supplements existing review workflows by providing a global view of how autonomous actions unfold.

\begin{figure*}[!ht]
    \centering
    \ifx\includesvg\undefined
        \includegraphics[width=\textwidth]{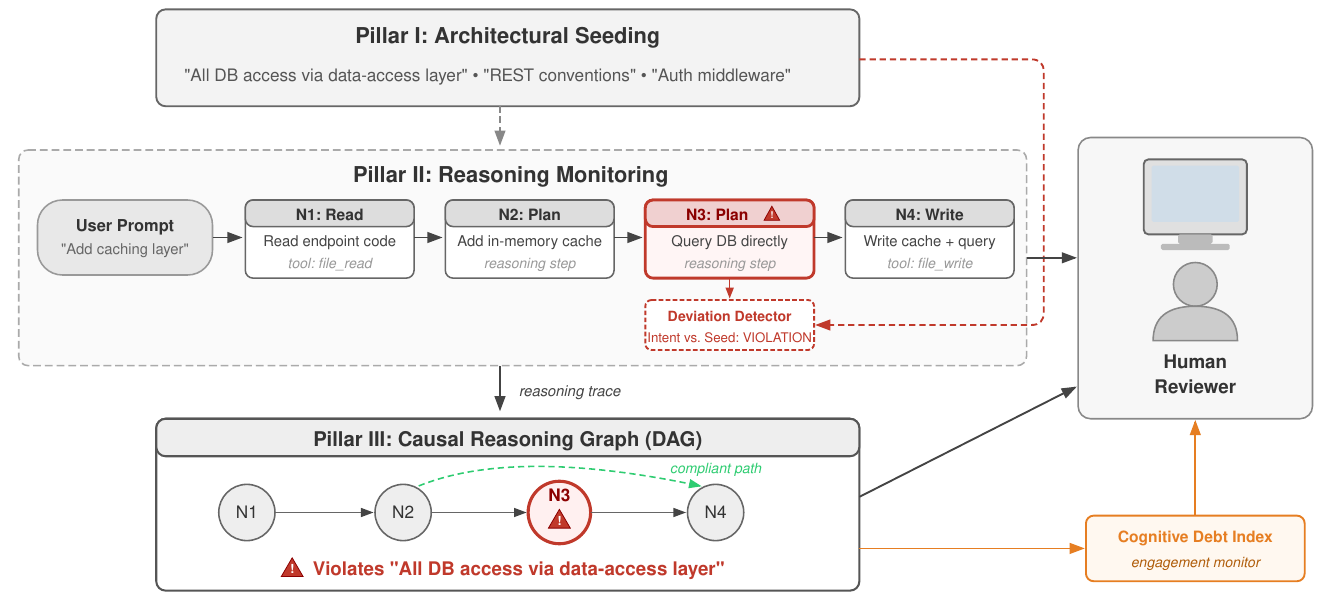}
    \else
        \includesvg[width=\textwidth]{figures/poe_pipeline_diagram.svg}
    \fi
    \caption{The Process-oriented Explainability (PoE) framework applied to a data‑access caching scenario. The figure illustrates the three pillars: Architectural Seeding, Reasoning Monitoring, and the Causal Reasoning Graph. It highlights how a deviation is detected when a reasoning step directly queries the database, violating the architectural rule. A cognitive debt index monitors reviewer engagement with the causal chain, signalling when human comprehension risks falling below the level required for substantive oversight.}
    \ifx\Description\undefined
    \else
        \Description{Diagram of the PoE framework with three pillars, showing reasoning steps N1–N4, a detected violation at N3, and an icon representing a human reviewer.}
    \fi
    \label{fig:poe-pipeline}
\end{figure*}

\section{Conclusion and Future Work}
\label{section:conclusion}
The HCXAI community must expand beyond localized transparency to address the broader challenge of agentic accountability. This work shows that without effective mitigation, the productivity gains of autonomous agents can be outweighed by the long-term costs of agentic entropy. By shifting attention from execution-level stack traces to intent-level reasoning traces, we offer a way to preserve architectural integrity as software systems evolve.
Critically, the risks we identify are not limited to the codebase itself. As warned by \citet{kosmyna2025your,lin2026}, sustained AI delegation erodes the user's ability to explain, verify, or intervene. In the context of agentic software development, this gap manifests as cognitive debt. Our PoE framework directly targets this dynamic. By making the agent's causal reasoning chain legible and auditable, it supports what \citet{lin2026} define as the `Cognitive Integrity Threshold' (CIT), i.e., the minimum level of comprehension required for oversight to remain substantive rather than performative. The cognitive debt index proposed in our implementation operationalizes this threshold in practice, flagging when reviewer engagement with the causal graph drops below a level sufficient to detect architectural drift. Future research should therefore treat cognitive drift as a first-class metric alongside code quality indicators, developing empirical benchmarks for CIT in agentic development contexts and studying how process-level interfaces can actively counteract comprehension decay over time.

Future research should prioritize the development of \emph{Standardized Agentic Traces}, representing a universal, schema-agnostic log format. Such a standard would facilitate interoperability across agents and human-centric audit tools, ensuring that reasoning remains transparent across diverse development environments. In addition, empirical studies are needed to quantify the reduction in cognitive load provided by process-level `cognitive debugging' interfaces compared to traditional code review practices \citep{Fakhoury2024}. Ultimately, the goal of agentic XAI is to shift the human–agent relationship from one of blind trust in velocity to one of shared, verifiable understanding of architectural intent, an understanding that must be actively maintained in both the codebase and the minds of the people who oversee it.

\section*{Acknowledgements}
This work was partly conducted within the framework of the EUonAIR Centre of Excellence in Responsible AI and Education. The authors have been supported by a grant from Movetia, which is funded by the Swiss Confederation.

\printbibliography 

\end{document}